# Extended topological mode in a one-dimensional non-Hermitian acoustic crystal

Xulong Wang, Wei Wang, Guancong Ma*

Department of Physics, Hong Kong Baptist University, Kowloon Tong, Hong Kong, China

*Email: phgcma@hkbu.edu.hk

**Abstract**

In Hermitian topological systems, topological modes (TMs) are bound to interfaces or defects of a lattice. Recent discoveries show that non-Hermitian effects can reshape the wavefunctions of the TMs and even turn them into extended modes occupying the entire bulk lattice. In this letter, we experimentally demonstrate such an extended TM (ETM) in a one-dimensional (1D) non-Hermitian acoustic topological crystal. The acoustic crystal is formed by a serie of coupled acoustic resonant cavities, and the non-Hermiticity is introduced as the non-reciprocal coupling coefficient using active electroacoustic controllers (AECs). Our work highlights the potential universality of ETMs in different physical systems and resolves the technical challenges in the further study of ETMs in acoustic waves.

Hermitian topological matters sustain topological modes (TMs) that are bound states localized at interfaces, edges, or defects of a specimen[1–3]. The TMs possess properties such as robustness against perturbations, which are highly desirable for many applications. The studies of topological matters have expanded beyond condensed matter physics and are nowadays pursued in photonics[4], acoustics, and mechanics[5,6]. In all these systems, the existence of TMs rests on the non-trivial bulk-band topology, which implies that a bulk lattice is indispensable. This requirement entails high costs, large footprints, and uneconomic use of space and materials for topological devices.

Recent studies of non-Hermitian physics have unveiled a range of new physical phenomena[7–10], such as non-Hermitian skin effect (NHSE)[11–13]. NHSE describes the appearance of a large number of skin modes, which are localized bulk eigenstates in open-boundary lattices. This is in stark contrast to the extended Bloch waves observed in Hermitian systems. Recently, it has been found that NHSE can conversely affect the wavefunctions of TMs, and can even turn them into completely extended states occupying every unit cell in the entire lattice[14–17]. Such extended

TMs (ETMs) have been experimentally demonstrated in 1D and two-dimension (2D) non-Hermitian mechanical lattices[16,17]. Because the theoretical models of ETMs are based on tight-binding theories that are universal, we anticipate similar phenomena to emerge in acoustics, electromagnetism, and photonics, etc. However, so far ETMs have not been observed in all these areas. Here, we demonstrate an acoustic ETM in a non-Hermitian Su-Schrieffer-Heeger (NH-SSH) lattice formed by coupled acoustic cavities[5,6,18] with active electroacoustic controllers (AECs) controlled by acoustic feedbacks[19–21] generating non-reciprocal coupling. Our work contributes to non-Hermitian acoustics and topological acoustics and opens new ways to manipulate acoustic waves.

Our experiment is based on a 1D NH-SSH chain with non-reciprocal hopping, as depicted in Fig. 1a. The Hamiltonian of a finite lattice under open boundary condition (OBC) is

$$H = \sum_{i=1}^{N-1}\left[v a_i^\dagger b_i + w a_{i+1}^\dagger b_i + (v+\delta) a_i b_i^\dagger + w a_{i+1} b_i^\dagger\right], \tag{1}$$

where $a_i^\dagger$ and $a_i$ ($b_i^\dagger$ and $b_i$) are the creation and annihilation operators of $A_i$ site ($B_i$ site) in unit cell $i$, $N = 15$ is the number of unit cells, and $v = -70.4\ \mathrm{rad/s}$ and $w = -217.4\ \mathrm{rad/s}$ are the intracell and intercell hopping, respectively. These values are retrieved from the experimental system. $\delta$ represents the non-reciprocal hopping term from the site $A_i$ to $B_i$. In the Hermitian case, i.e., $\delta = 0$, the lattice is topological and the two bulk bands (Fig. 1b) are characterized by a quantized Zak phase of $\pi$[5,22]. Because we keep only one site in the last unit cell (site $A_N$), there is only one TM at zero energy localized at the left boundary, as shown in Fig. 1b, d. The bulk modes are extended states, which can be seen by the wavefunction average, defined as $\bar{\Psi}(x) \equiv \frac{1}{2N-1}\sum_{j=1}^{2N-1}\left|\psi_j(x)\right|^2$ with $\psi_j(x)$ being the eigenfunction of the $j$th mode. (Our system has $2N-1$ sites in total, there are $2N-1$ modes.) In Fig. 1d, it is seen that $\bar{\Psi}(x)$ is uniformly distributed over the entire lattice. When $\delta \neq 0$, the system becomes non-Hermitian, and its spectrum under periodic boundary condition (PBC) forms two loops in the complex plane (Fig. 1c), signifying the existence of NHSE[23]. This can be seen in Fig. 1e, where $\delta = w - v$. $\bar{\Psi}(x)$ clearly collapses towards the right side of the lattice under OBC. Meanwhile, the TM is turned into a fully extended mode because $\delta$ counters the spatial exponential decay that characterizes the Hermitian TM. This is the ETM that we aim to demonstrate in acoustics.

The experimental platform is an acoustic crystal consisting of 9 cuboid cavities ($35 \times 35 \times 80\ \mathrm{mm}^3$ in length, width, and height, respectively) connected by channels 40 mm in length with periodically alternating cross-sectional areas ($23.7\ \mathrm{mm}^2$ and $165\ \mathrm{mm}^2$ for the intracell hopping $v$ and intercell hopping $w$, respectively), as shown in Fig. 2a. To observe the TM, we place a loudspeaker at cavity $A_1$ and measure spectral responses of every cavity. Figure 2b shows the normalized pressure responses at cavity $B_3$ and $A_2$, respectively. Because the TM only has support in sublattice $A$, two bulk bands separated by a bandgap are seen in the results measured at cavity $B_3$. The TM is observed as the mid-gap response peak at 2145 Hz in the results measured at cavity $A_2$.

The next step is to add non-reciprocal coupling to the acoustic system. This is achieved by using AECs, as shown in Fig. 3a, b. The device consists of a microphone (Panasonic WM-G10DT502), a loudspeaker and a custom-made printed circuit board (PCB) that integrates a phase shifter, an amplifier, and a microcontroller (MCU, ESP32 LoLin32 Lite). The MCU is programmed to function as two independent digital potentiometers that control the phase shifter and the amplifier, respectively. The phase shifter compensates for the phase delay induced by other electric components. This is to ensure that $\delta$ is real. The amplifier determines the strength of the non-reciprocal coupling, $|\delta|$. To verify the successful realization of acoustic non-reciprocity and to benchmark the AEC's performance, we use a simple two-level model

$$H_2 = (\omega_0 - i\gamma)\mathrm{I}_2 + \begin{pmatrix} 0 & \kappa_0 \\ \kappa_0 + \delta & 0 \end{pmatrix}, \tag{2}$$

wherein $\mathrm{I}_2$ is a $2 \times 2$ identify matrix, $\omega_0$ is the resonant frequency of the cavity, and $\gamma$ represents the dissipative rate of the cavities, $\kappa_0$ denotes the reciprocal part of the coupling. The eigenvalues of $H_2$ are $\omega_{1,2} = \omega_0 - i\gamma \pm \sqrt{\kappa_0(\kappa_0 + \delta)}$, which reach an exceptional point at $\delta/\kappa_0 = -1$. The acoustic system for Eq. (2) is simply two coupled cavities, denoted $C_1$ and $C_2$, as shown in Fig. 3a. We excite at cavity $C_1$ with a loudspeaker and measure the response function at $C_2$ at different non-reciprocal coupling generated by an AEC. The results are shown in Fig. 3c as markers which agree well with the responses computed using Green's function (see, e.g., refs.[16,24] for more details of the method). Here, the retrieved parameters are onsite resonant frequency $f_0 = \frac{\omega_0}{2\pi} = 2145\ \mathrm{Hz}$, dissipative rate $\gamma = 70.4\ \mathrm{rad/s}$, and reciprocal coupling strength $\kappa_0 = -55.3\ \mathrm{rad/s}$, respectively. In Fig. 3d, the retrieved $\mathrm{Re}(\omega_{1,2})$ are plotted as functions of $\delta/\kappa_0$, in which the exceptional point is clearly identified at $\delta/\kappa_0 = -1$. These experimental results confirm that our AECs can indeed realize non-reciprocal hopping.

The AECs are then combined with the acoustic crystal, as shown in Fig. 2a. We then use a loudspeaker to excite the system from cavity $A_1$ at 2145 Hz, which is the eigenfrequency of the TM, and measure the pressure across the entire acoustic crystal for different non-reciprocal hopping.

The results are shown in Fig. 2c. We observe that the pressure responses in the bulk increase with $|\delta|$. And when $\delta = -147.0\ \mathrm{rad/s}$, the response profile is almost flat while maintaining the bipartite characteristics, which is evidence of the emergence of ETM. The same pressure distributions are obtained in numerical simulation using finite-element solver COMSOL Multiphysics, as shown in Fig. 2d. Both the experimental and simulation results agree well with theoretical predictions based on tight-binding model [Eq. (1)].

In conclusion, we report the realization and observation of an ETM in acoustics for the first time, wherein the non-reciprocal hopping is realized using the AECs. Together with the previously demonstration in mechanical lattices[16,17], our results show that the ETM is, in principle, a universal non-Hermitian phenomenon, which should also appear in other realms, such as electromagnetism, photonics, and so on. We envision ETM to have good application potential, and to be particularly valuable for large-area single-mode devices.

**Acknowledgements.** This work is supported by the Hong Kong Research Grants Council (RFS2223-2S01, 12301822, 12302420, 12302419) and the Ministry of Science and Technology of China (2022YFA1404400).

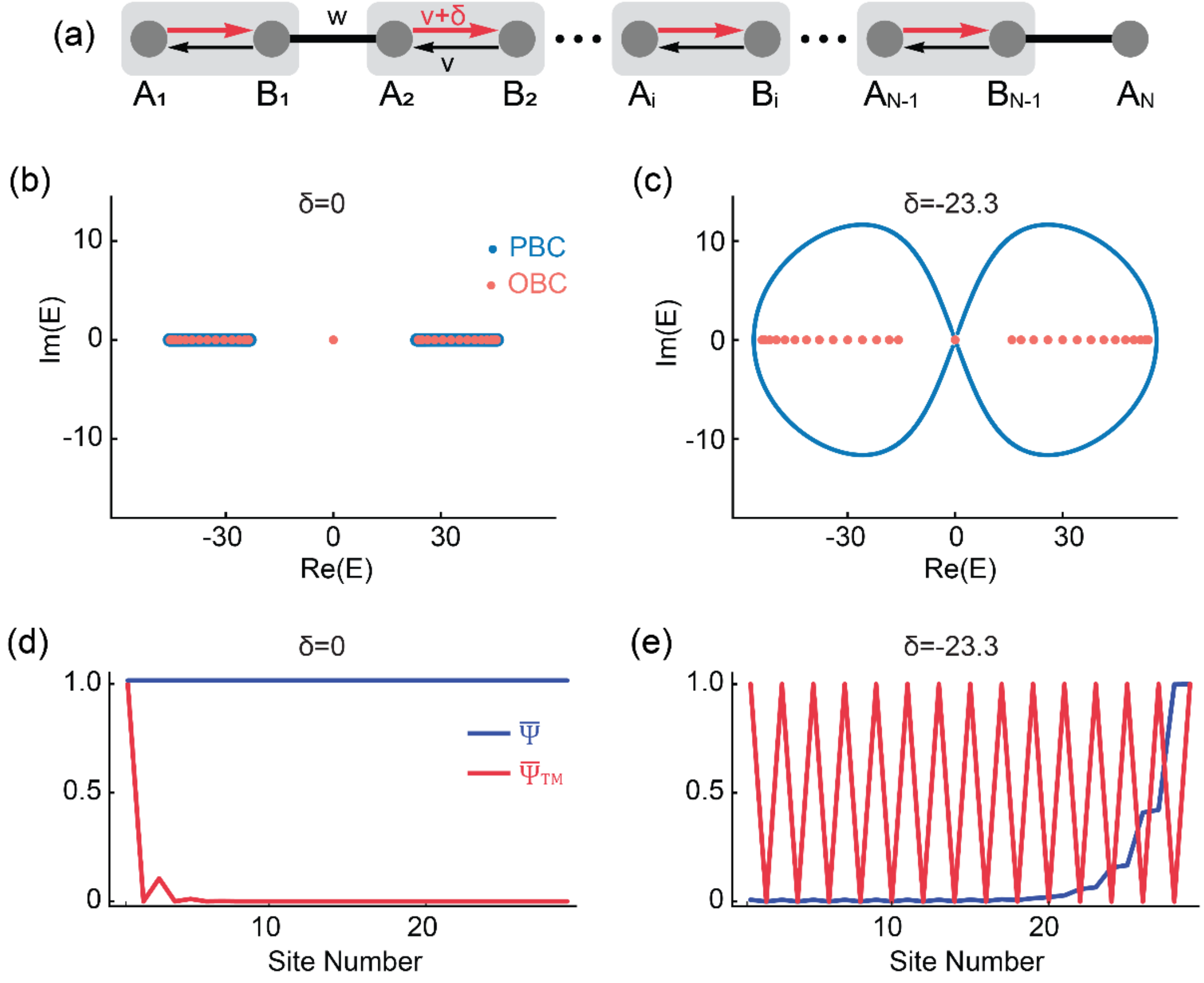

Fig. 1 (a) A 1D NH-SSH chain with non-reciprocal intracell hopping $v$ (from $B_{N-1}$ to $A_{N-1}$) and $v+\delta$ (from $A_{N-1}$ to $B_{N-1}$) under OBC. It has 29 sites in total. (b) shows the PBC and OBC spectra of a Hermitian SSH chain, i.e., $\delta = 0$, and (c) shows the complex energy spectra of the NH-SSH chain under PBC and OBC with $\delta = w - v = -147$ rad/s. (d) and (e) respectively show the wavefunction of the TM ($\overline{\Psi}_{\mathrm{TM}}(x) = |\psi_{\mathrm{TM}}(x)|^2$, red) and the wavefunction average of all states [$\overline{\Psi}(x)$, blue, definition in the main text]. (d) In the Hermitian case ($\delta = 0$), the TM is localized at the left boundary of the chain and the bulk modes are extended. (e) When $\delta = w - v$, NH skin modes and ETM are seen.

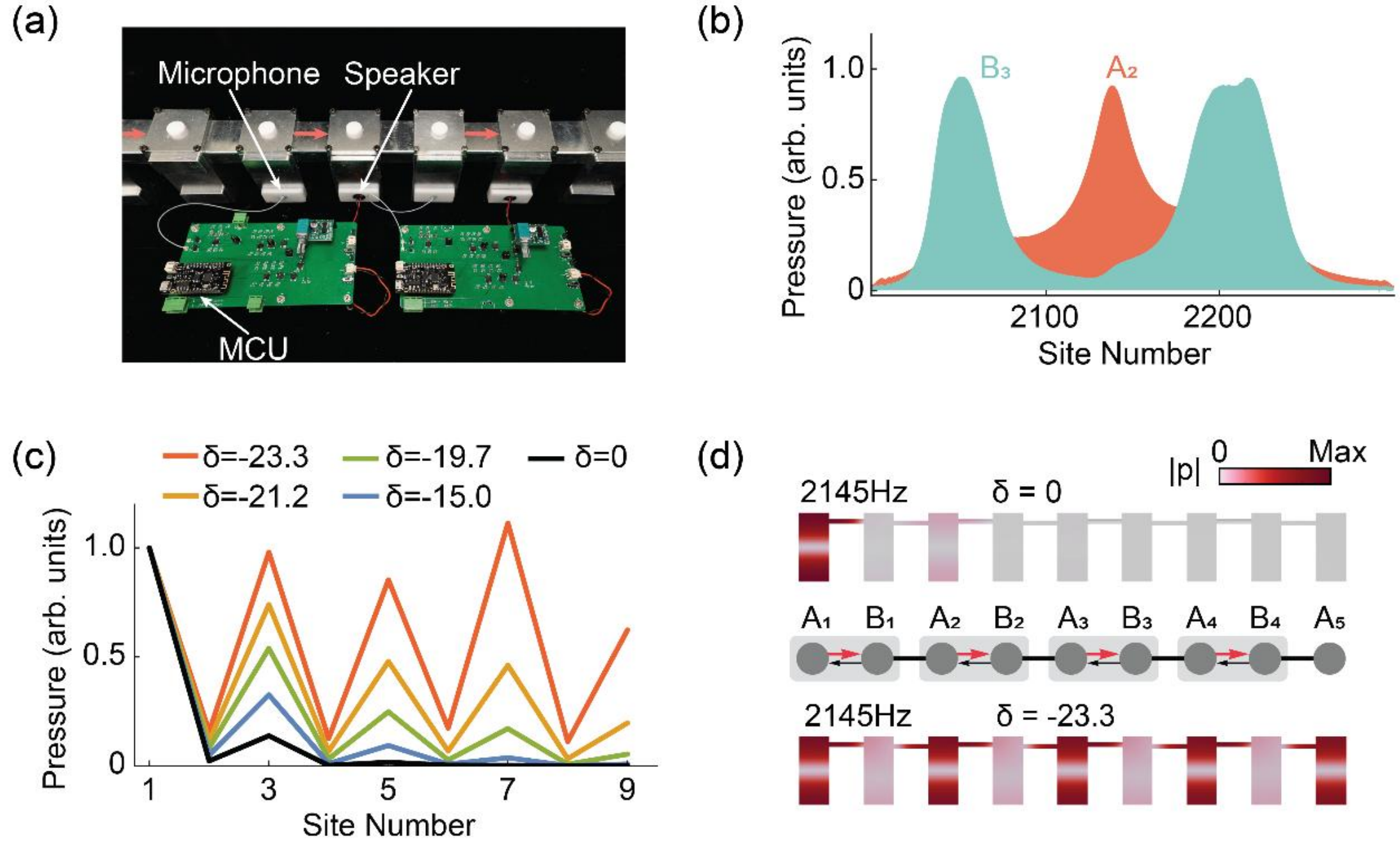

Fig. 2 (a) The acoustic crystal that realizes the 1D NH-SSH model. The non-reciprocal coupling between cavities (indicated by red arrows) is imposed by the AECs containing a microphone, a speaker at the bottom of cavities and an MCU. (b) The normalized spectral responses measured at acoustic cavity $B_3$ and $A_2$, respectively, with excitation at $A_1$. (c) The measured pressure responses (normalized amplitudes) at 2145 Hz under different $\delta$. The ETM is observed at $\delta = -147$ rad/s. (d) Numerically simulated pressure eigenfunctions at two different $\delta$.

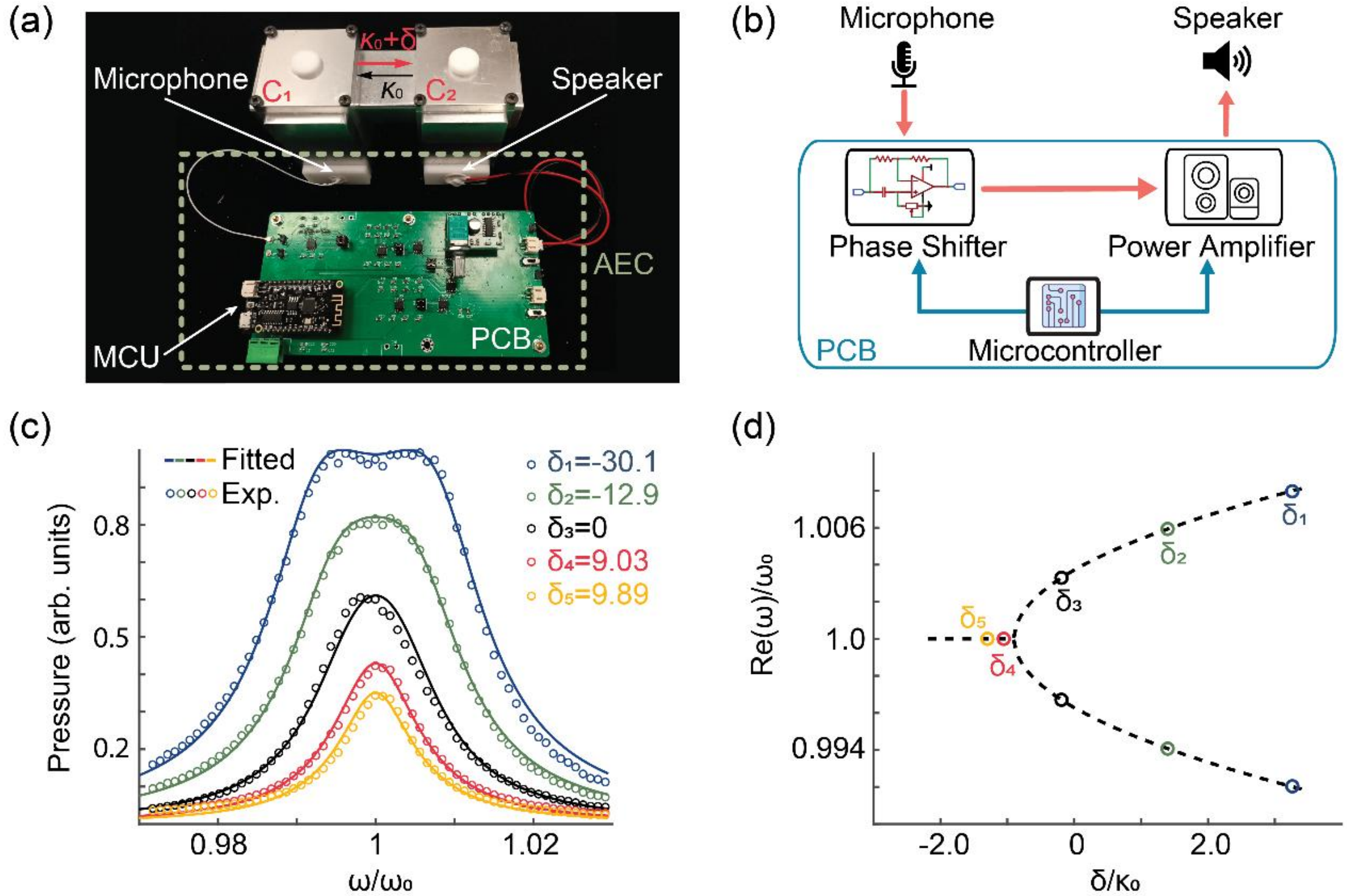

Fig. 3 (a) A photo showing two coupled cavities, denoted $C_1$ and $C_2$, and an AEC. (b) The schematic drawing of the AEC. (c) The normalized pressure responses measured in cavities $C_2$ at different non-reciprocal coupling $\delta$. The source is placed at cavity $C_1$. (d) The real parts of eigenfrequencies as a function of $\delta/\kappa_0$. The dashed curves are theoretical results obtained using Model (2). An exceptional point is identified at $\delta/\kappa_0 = -1$.